\documentclass[journal]{IEEEtran} % use the `journal` option for ITherm conference style
\IEEEoverridecommandlockouts
\usepackage{cite}
\usepackage{amsmath,amssymb,amsfonts}
\usepackage{algorithmic}
\usepackage{graphicx}
\usepackage{booktabs}
\usepackage{url}
\usepackage{textcomp}
\usepackage{xcolor}
\def\BibTeX{{\rm B\kern-.05em{\sc i\kern-.025em b}\kern-.08em
    T\kern-.1667em\lower.7ex\hbox{E}\kern-.125emX}}

\usepackage{tikz}
\usetikzlibrary{arrows,decorations,shapes,backgrounds}
\usepackage{pgfplots}
\pgfplotsset{compat=newest}
\usepackage{stfloats}
\usepackage{mathtools}

\DeclarePairedDelimiter\floor{\lfloor}{\rfloor}

\def\Lon{{M_{\rm ZC}}}
\def\Nzc{{N_{\rm ZC}}}

\def\Prsrp{P_{\mathrm{LP-RSRP}}}
\def\Prssi{P_{\mathrm{LP-RSSI}}}
\def\Rrsrq{R_{\mathrm{LP-RSRQ}}}

\begin{document}

\title{Low-Power Wake-Up Signal Design in 3GPP 5G-Advanced Release 19\\
% delete or comment-out the following line before submission
% {\footnotesize \textsuperscript{*}Note: Sub-titles are not captured in Xplore and should not be used}
%\thanks{This project has been supported by the French FRAMExG program.}
}

\author{%%%% author names
    \IEEEauthorblockN{Sebastian Wagner}% first author
    %\IEEEauthorblockN{1\textsuperscript{st} Sebastian Wagner}% first author
    %, \IEEEauthorblockN{2\textsuperscript{nd} Kien Le Trung}% delete this line if not needed
    %, \IEEEauthorblockN{2\textsuperscript{rd} Raymond Knopp}% delete this line if not needed
    % duplicate the line above as many times as needed to list all authors
    \\%%%% author affiliations
    \IEEEauthorblockA{\textit{EURECOM, Sophia-Antipolis, France}\\
    sebastian.wagner@eurecom.fr}\\% first affiliation
    % \IEEEauthorblockA{\textit{dept. name of organization (of Aff.), City, Country if needed}}\\% delete this line if not needed
    % duplicate the line above as many times as needed to list all affiliations
    %%%% corresponding author contact details
    %\IEEEauthorblockA{sebastian.wagner@eurecom.fr}
}

\maketitle

\begin{abstract}
    The Low-Power Wake-Up Signal (LP-WUS) and Low-Power Synchronization Signal (LP-SS), introduced in 3GPP 5G-Advanced Release 19, mark an important advancement toward power-efficient IoT communications. This paper provides a comprehensive overview of the LP-WUS procedures in the RRC\_IDLE and RRC\_INACTIVE states and summarizes the key physical-layer design aspects. The LP-WUS is intended to be detected by a low-power energy detector (ED), allowing the main radio (MR) to remain switched off, thereby enabling substantial power savings compared to conventional 5G paging mechanisms. As such, LP-WUS is considered the baseline for 6G WUS design. Furthermore, different receiver architectures are evaluated, highlighting the inherent trade-offs between power-saving gains and coverage performance.
\end{abstract}


\section{Introduction}
Low-Power Wake Up Signals (LP-WUS) are essential in many low-power communication protocols such as LoRa, Bluetooth, or WiFi \cite{b2}. These signals allow for the design and implementation of low-power radios and thus contribute significantly to a reduction in power consumption for devices with application in (Industrial) Internet of Things (IoT). 

In cellular networks, 3GPP has conducted a Study Item (SI) on LP-WUS in Release 18, cf. \cite{b3}, \cite{wagnerWusR18}, with the goal to evaluate the potential reduction in power consumption of a 5G device equipped with a LP Wake-Up Radio (LR). Typically, a 5G device consumes tens of milliwatts even if it is not transmitting or receiving any data, a state called RRC\_IDLE/INACTIVE. This idle power consumption is due to the fact that the 5G device has to carry out periodic measurements (to support mobility) and check for potential paging messages (to be reachable), once per Discontinuous Reception (DRX) cycle. Hence, 5G IoT devices with long battery life on a single charge have long been elusive. 

Battery life is directly related to the DRX cycle. A longer DRX cycle results in improved battery life because the UE can stay in the low power sleep state. On the other hand, since the UE is not available during the DRX period, a longer DRX cycle increases the delay with which the UE can be reached by the network. Many network services have strict requirements on the delay with which the UE can be reached. Hence, the battery-life is inherently limited by the requirements of the network services. Examples of such services or use cases are actuators that operate sprinklers, the sprinklers have to be activated as soon as the smoke sensors detect a potential fire. Delays of tens of seconds are not acceptable.

The LP-WUS feature takes a step forward and enables significant improvements in power consumption in the idle state while maintaining latency requirements. To achieve this, two novel signals are introduced, the LP-WUS triggering the decoding of the paging message and the LP-SS to perform measurements. Both signals are processed by the LR, operating \textit{independently} of the 5G Main Radio (MR). In this way, the MR can remain powered down while the LP-WUR continues to search for potential LP-WUS and carry out measurements via LP-SS, thereby reducing energy consumption. This differentiates LP-WUS from existing power-saving features such as the 5G Rel-17 Paging-Early Indication (PEI) which is based on the Physical Data Control CHannel (PDCCH) and requires the MR for reception.

LP-WUS is also specified for the RRC\_CONNECTED mode. Similar to RRC\_IDLE/INACTIVE, the LP-WUS triggers PDCCH monitoring by the MR. However, since DL traffic arrival is significantly more likely in RRC\_CONNECTED, the MR cannot power down into an ultra-deep sleep state since it would take too long to power the MR up again. That is the main reason the power-saving gains are more pronounced in RRC\_IDLE/INACTIVE since the MR consumes almost no power in the ultra-deep sleep state. Consequently, this paper focuses on the LP-WUS operation in RRC\_IDLE/INACTIVE mode.

The paper is organized as follows: Section \ref{sec:operation} describes the LP-WUS procedures and configurations in RRC\_IDLE/INACTIVE mode. In Section \ref{sec:phy} the physical layer design of both LP-WUS and LP-SS is explained. Finally, Section \ref{sec:conclusion} provides a summary and conclusion of this contribution.

\section{LP-WUS Operation in RRC\_IDLE/INACTIVE}
\label{sec:operation}

In this section, we explain the overarching concepts and procedures when the UE is in RRC\_IDLE/INACTIVE mode before going into the details on physical layer design.

% - in-band operation: WUS operates withing NR DL band (ED more appropriate for single band operation)
% LR and MR can operate in different bands

The LP-WUS is configured via Radio Resource Control (RRC) signaling. In particular, the configuration \texttt{lowPower-Config-r19} is signaled in SIB1 within the \texttt{DownlinkConfigCommonSIB} information element, \cite{cr38331}.

During RRC\_IDLE/INACTIVE state, the UE does not have data to transmit and is not synchronized to the network. It has to follow two procedures: (i) decode the paging message to check for incoming traffic and (ii) carry out measurements for Radio Resource Management (RRM). These procedures are carried out periodically and the period is referred to DRX cycle $T$ spanning $320 ms$ to $2560 ms$.

In order to maintain the MR in an ultra low-power sleep state, these two procedures can be carried out with the LR by motoring two specifically designed low-power signals: (i) LP-WUS to trigger the paging reception and (ii) LP-SS to carry out measurements and synchronize the LR for LP-WUS decoding.

\subsection{Subgrouping for LP-WUS}
Every UE is assigned to a subgroup (SG) that can be signaled by the LP-WUS in the physical layer. There are two methods for subgrouping: (i) CN-assigned subgrouping and (ii) UE\_ID-based subgrouping, \cite{cr38304}. In CN-assigned subgrouping, the subgroup index $i_{SG}$ is provided by the AMF through NAS signaling. For UE\_ID-based subgrouping, the SG index is determined based on the UE\_ID, paging configuration and the number of SGs per Paging Occasion (PO).

A UE has to monitor the LP-WUS associated with its own SG as well as the LP-WUS associated to \textit{all} SGs withing of the PO.

\begin{figure*}[t]
    \centerline{\includegraphics[width=1.0\textwidth]{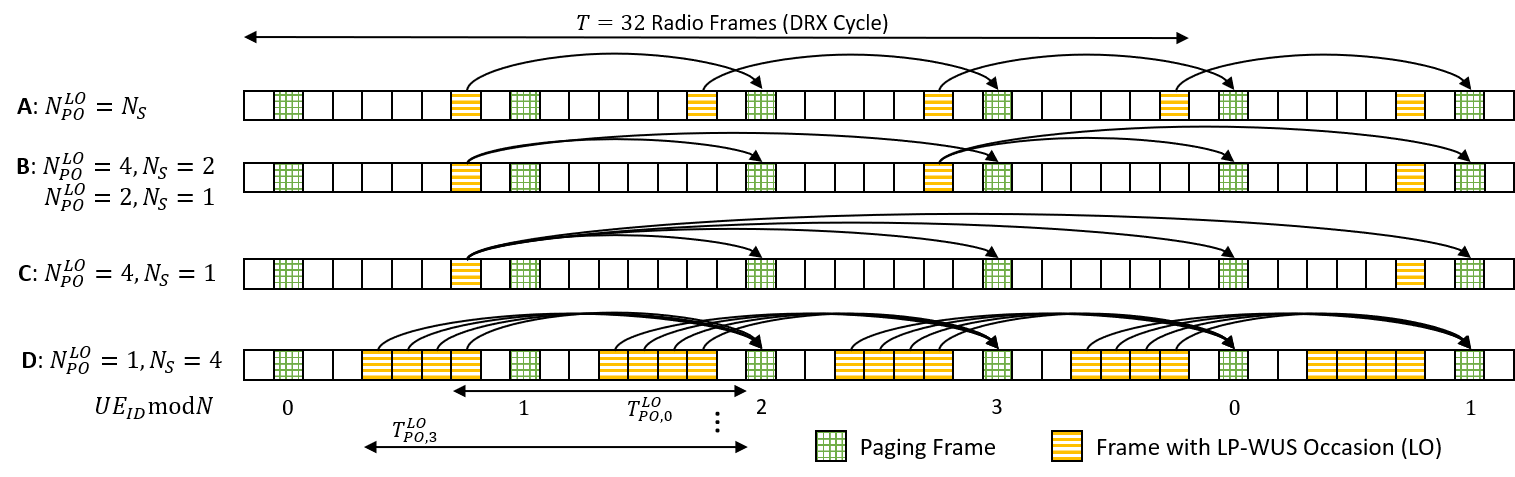}}
    \caption{Example of possible PO-to-LO associations with $T=32$ and $N=4$.}
    \label{fig:lo-po-mapping}
\end{figure*}

\subsection{LP-WUS Occasions and Paging Association}
The LP-WUS is transmitted during an LP-WUS Occasion (LO) and can be associated to $N_{PO}^{LO}\in\{1,2,4\}$ POs. The index $i_{PO}$ of the PO associated with an LO is given by, \cite{ts38213}
\begin{equation}
    i_{PO} = \left[(UE_{ID}\mod{N}) N_S + i_S\right]~\mathrm{mod}~{N_{PO}^{LO}},
\end{equation}
where $UE_{ID}$ is a 16-bit UE identity, $N$ is the number of paging frames per DRX cycle, $N_S$ is the number POs per Paging Frame (PF) and $i_S$ is the index of the PO within the PF. After the UE computes its $i_{PO}$, the time location of the corresponding is LO determined by a time offset $T_{PO}^{LO}$ to a reference PF (RPF) provided in SIB1. The System Frame Number (SFN) of the reference PF is calculated as \cite{cr38304}
\begin{equation}
    SFN_{RPF} = SFN_{PF} - \floor*{\frac{i_{PO}}{N_S}} \frac{T}{N}
\end{equation}
where $SFN_{PF}$ is the SFN of the PF.

An example of possible PO-to-LO associations is shown in Figure \ref{fig:lo-po-mapping}. Depending on the configuration, a LO can be associated with 1, 2 or 4 PFs. The configuration $N_S > N_{PO}^{LO}$ (Case D) is supported by indicating multiple values of $T_{PO}^{LO}$ in SIB1, e.g. if $N_S=4$ and $N_{PO}^{LO}=1$ there are 4 time offsets signaled because each of the four POs in the PF has its own LO. The network configures $T_{PO}^{LO}$ depending on the wake-up delay of the MR reported by the UEs and the SSB periodicity. For instance, if the UEs with the same PO report a wake-up delay of $70ms$, the network configures $T_{PO}^{LO}\geq 70ms$ s.t. there are at least three SSBs available between the LO and PO for measurements and synchronization to receive the paging message with the MR.

Smaller values of $N_{PO}^{LO}$ support more SGs per PO and each LO can address all SGs in the associated POs which is described in more detail in the next section.

\begin{figure*}[t]
\includegraphics[width=1.0\textwidth]{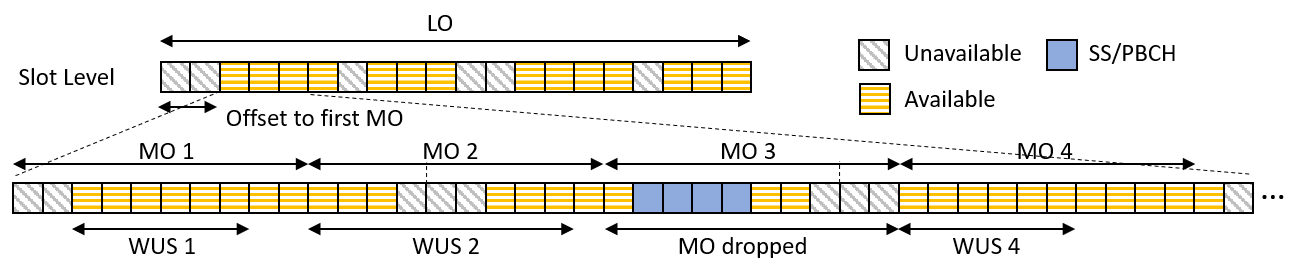}
\caption{Example of MO configuration with $L_{MO}=10$ and $L=6$ OFDM symbols. LO contains 20 slots.}
\label{fig:mo-config}
\end{figure*}

\subsection{Subgroup Signaling in LO}

The LP-WUS contains a maximum of $B_{max}=5$ information bits and therefore can, at most, signal 32 values, called \textit{codepoints}. It is important to retain the possibility to wake up all groups and hence the maximum number of subgroups is 31. Since one LO can be associated to $N_{PO}^{LO}$ POs, the \textit{maximum} number of subgroups per PO is $N_{SG,max}^{PO}\in\{31,15,7\}$ for $N_{PO}^{LO}\in\{1,2,4\}$, respectively. The number of configured subgroups $N_{SG}^{PO}$ is signaled in SIB1.

The codepoints $c_{SG}$ for each SG index $i_{SG}$ depend on the $i_{PO}$ and $N_{SG}^{PO}$ as
\begin{equation}
c_{SG} = 
    \begin{cases}
        i_{PO} & \text{if $N_{SG}^{PO}=1$} \\
        (i_{PO} + 1) + i_{SG} & \text{if $N_{SG}^{PO}>1$} ,
    \end{cases}
\end{equation}
and the the codepoint $c_{all}$ for all SGs $c_{all}=(i_{PO} + 1)(N_{SG}^{PO} + 1)-1$. An example is provided in Table \ref{tab:codepoints} where the maximum subgrouping is assumed. As described previously, the various $i_{PO}$ may correspond to different PFs.

\subsection{LP-WUS Occasions within an LO}

Each LO can contain $N_{LO}^{MO}\in\{1,2,3,4\}$ LP-WUS Monitoring Occasions (MO) per beam. As described above, it is important to understand that a PO is associated to an LO and \textit{not} to a MO. By configuring multiple MOs, the network can wake up multiple SGs associated with the LO, because the UE has to monitor \textit{all} MOs within its LO. For example, consider case C in Figure \ref{fig:lo-po-mapping}, assuming $N_{LO}^{MO}=4$, the network can either address SGs in each PO or 4 SGs in a single PO, significantly enhancing paging flexibility. 

\begin{table}[ht!]
  \centering
  \begin{tabular}{llll}
    \toprule
    \textbf{$N_{SG}^{PO}$} & $i_{PO}$ & $c_{SG}$ & $c_{all}$ \\
    \midrule
    1 & 0 & $\{0,1,...,30\}$    & $\{31\}$ \\
    2 & 0 & $\{0,1,..,14\}$     & $\{15\}$ \\
      & 1 & $\{16,17,...,30\}$  & $\{31\}$ \\
    4 & 0 & $\{0,1,...,6\}$     & $\{7\}$ \\
      & 1 & $\{8,9,...,14\}$    & $\{15\}$ \\
      & 2 & $\{16,17,...,22\}$  & $\{23\}$ \\
      & 3 & $\{24,25,...,30\}$  & $\{31\}$ \\
    \bottomrule
  \end{tabular}
  \vspace{5pt}
  \caption{Example of codepoints for $N_S = N_{SG}^{PO}={1,2,4}$ and $N_{SG}^{PO} = N_{SG,max}^{PO}$.}
  \label{tab:codepoints}
\end{table}

%\vspace{-10pt}

The configured number MOs are associated to beams. Either to the same beams as SS/PBCH indicated in SIB1 or to a subset of those beams signaled in the LP-WUS configuration.

The duration of a MO, in number of OFDM symbols, is indicated by \textit{two} values: (i) a \textit{nominal} MO duration $L_{MO}$ and (ii) an \textit{actual} MO duration $L$. The difference between the two stems from the number OFDM symbols that are \textit{available} for LP-WUS transmission. $L$ is the configured duration of the LP-WUS and $L_{MO}$ the duration of the MO in which the LP-WUS is transmitted. If there are not enough OFDM symbols available in the MO, i.e. $L > L_{MO}$, the UE skips this MO, \cite{ts38213}.

An example configuration is shown in Figure \ref{fig:mo-config}, where a LO frame consists of 20 slots ($30kHz$ subcarrier spacing). The start of the first MO within the LO is indicated by an offset value, which is 28 OFDM symbols (2 slots) in this example. Both, the available slots and OFDM symbols are configured via a bitmap with periodicity of 10 slots and the bitmap for the available OFDM symbols is 14 bits. The first to symbols are unavailable for LP-WUS, e.g. used for PDCCH, as well as the last symbol, e.g. UL transmission. The figure shows the first 4 MOs which can correspond to a single beam, 2 beams or 4 beams, depending on the network configuration. The MO and WUS durations are set to $L_{MO}=10$ and $L=6$ OFDM symbols, respectively. Notice that the actual WUS has different durations spanning 6 or 9 OFDM symbols. Also, the gap between the different WUS transmissions can vary. The third MO is dropped in this example because there are not enough symbols available for the WUS due to an SS/PBCH transmission.

\subsection{LP Measurement Procedure}
To allow for efficient Radio Resource Management (RRM), the UE has to carry out measurements for cell (re)selection, mobility procedures, beam management etc. Those measurements, e.g. SS-RSRP, SS-RSRQ, etc. are obtained by the MR from the synchronization signal block (SSB). 

In order to keep the MR in ultra-deep sleep, the network configures periodic LP-SS which can be used by the LR to (i) carry out the required RRM measurements and (ii) obtain (coarse) time-frequency synchronization to receive the LP-WUS and subsequent paging messages.

The LP-SS is a periodic signal with a configurable periodicity of $320ms$ or $160ms$. Those values trade-off resource overhead and synchronization accuracy. A lower periodicity increases resource overhead but allows to compensate for larger time and frequency drifts.

\begin{figure*}[t]
    \centerline{\includegraphics[width=1\textwidth]{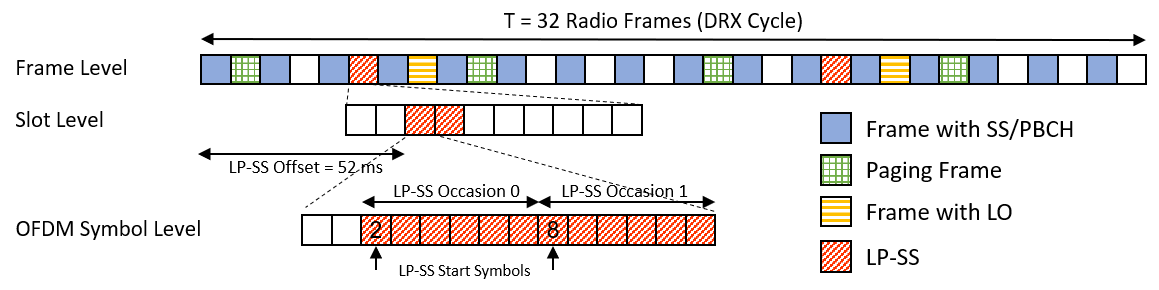}}
    \caption{Example of LP-SS configuration with periodicity of $160ms$ and 4 configured beams.}
    \label{fig:lpssMapping}
\end{figure*}

The LP-SS spans $L_{LPSS}\in\{4,6,8\}$ \textit{consecutive} OFDM symbols \textit{within} a slot and there is one LP-SS per configured beam. The LP-SS is configured with an offset (in $ms$) w.r.t. SFN0 and one or two start symbols. Figure \ref{fig:lpssMapping} shows an example configuration with $L_{LPSS}=6$ and two LP-SS occasions per slot each corresponding to one beam.

% - max overlay sequences 16 (M=1) and 8 (M=2) and 4 (M=4) (R1 120)
%  (discussion of amount of correlation that need to be carried out)
% Maximum number of subgroups per PO is 31 (5 bits with one codepoint remaining to wake up everyone)

% Simililar power-saving functionalities such as LTE-WUS or PEI

% What are the things the UE has to do ?
% Procedures are defined in TS38.304 and TS38.331

%  - Picture of the procedure with deep sleep time, minium time that is configured
%  - also talk about LP-SS for radio-resource monitoring, LP-RSRP
%  - what information is carried in LP-WUS (Codepoint-concept)

%\subsection{RRC CONNECTED}
%IN RRC CONNECTED mode, the UE is synchronized to the network and monitors the PDCCH for incoming DL transmissions or UL grants. 

% - Codepoint-based representation (120)
% - maximum number of Codepoints per MO per UE is 8 (UE capability) (120)
% - LP-WUS SCS is same as the active DL-BWP (120)
% - LP-WUS in the same carrier as active DL-BWP but can be outside active DL-BWP
%   Basic capability LP-WUS within active DL-BWP

% Legacy power-saving techniques such as Rel-16 DCP
% MR is not in ultra-deep sleep
% RRM measurements are performed by MR
% Potential support of LR for RRM measurements ?

%  - Instead of PDCCH decode WUS which triggers PDCCH monitoring
%  - No sleep states as in RRC IDLE
%  - saves power because PDCCH monitoring is power-consuming
%  - there are already features that do this, e.g. PDCCH skipping etc
%  - what information is carried in LP-WUS

\section{Physical Layer Design}\label{sec:phy-design}
\label{sec:phy}
% - Introduction to OOK (Why it was specified in TD)
% - Frequency allocation, BW, GB, why 11PRBs for 15 and 30kHz
% - SCS for LP-SS and LP-WUS
% - FR1 and FR2
% - 2 signals, LP-SS and LP-WUS
% - Both follow the same waveform generation but LP-SS are specified sequences

%\subsection{Multi-Carrier On-Off Keying}

The underlying modulation is based on On-Off Keying (OOK) allowing for a low-power receiver implementation, i.e. envelope/energy detection. It is a special case of Amplitude Shift Keying (ASK) with only two amplitudes, ON and OFF. Whenever the input bit $b=1$, the transmitter sends an ON-signal $\mathbf{r}=[r_0,r_1,...,r_{\Lon-1}]$ of length $\Lon$ and an OFF-signal $\mathbf{r}=\mathbf{0}$ of equal length if $b=0$. We refer to the ON and OFF signals as OOK symbols.

When applied to a multi-carrier system, such as OFDM, OOK is also referred to as multi-carrier (MC) OOK, because, in frequency domain, the OOK symbols typically span multiple sub-carriers.

One design criteria for LP-WUS is its seamless integration in the existing 5G gNB transmission chain. The Rel-19 LP-WUS and LP-SS support $M\in\{1,2,4\}$ OOK symbols per OFDM symbol which are generated in time-domain and subsequently transformed via DFT into the frequency domain to be mapped to the existing 5G OFDM resource grid. For a fixed LP-WUS signal bandwidth (excluding guard-bands) of $N_{SC}^{WUS}=132$ subcarriers, the length of the OOK sequences are $\Lon=N_{SC}^{WUS}/M\in\{132,66,33\}$ for $M\in\{1,2,4\}$, respectively.

% - OOK time domain -> N-DFT -> N sub-carriers

% - illustration that shows ideal ON-OFF signal with the different timings for various values of M

% Explain the details of LP-WUS later

\subsection{LP-WUS Design}

Let $\mathbf{b}=[b_0,b_1,...,b_{B-1}]$ designate the information bits (codepoint) of length $B$ carried by the WUS, where $B\leq 5$ bits. The corresponding time-domain signal $\mathbf{s} =[\mathbf{s}_{0},\mathbf{s}_{1},...,\mathbf{s}_{L-1}]$ of length $N_{SC}^{WUS}L$ spans $L$ OFDM symbols (not necessarily consecutive). Each of the OOK waveforms within an OFDM symbol $\mathbf{s}_{l}=[\mathbf{r}_0,\mathbf{r}_1,...,\mathbf{r}_{M-1}]$ is constituted of $M$ OOK symbols. Thus, $\mathbf{s}$, $\mathbf{s}_l$ and $\mathbf{S}_l$ denote the time-domain OOK signals of the entire WUS transmission, the WUS per OFDM symbol and the frequency-domain representation, respectively.

\begin{figure}[htbp]
    \centerline{\includegraphics[width=0.5\textwidth]{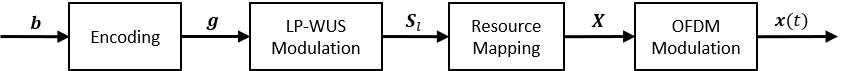}}
    \caption{Block-diagram of LP-WUS waveform generation.}
    \label{fig:wave_generation}
\end{figure}

A general block diagram of the WUS signal generation is shown in Figure \ref{fig:wave_generation}. The information bits $\mathbf{b}$ are encoded and the resulting $G$ coded bits $\mathbf{g}=[g_0,g_1,...,g_{G-1}]$ are modulated onto $L$ OFDM symbols each carrying $M$ coded bits. Subsequently, the WUS in frequency-domain $\mathbf{S}_l$ spanning $N_{SC}^{WUS}=132$ subcarriers is mapped to the overall resources $\mathbf{X}$ and OFDM-modulated onto $L$ OFDM symbols resulting in the time-domain signal $\mathbf{x}(t)$.

\subsubsection{Encoding}
The encoding procedure consists of three steps: (i) Channel coding, (ii) rate-matching (RM) followed by (iii) line coding. Channel coding is necessary to meet the requirements on False-Alarm Rate (FAR) and Missed-Detection Rate (MDR) and uses the coding schemes for small block lengths defined in \cite[Section 5.3.3]{ts38212} with $Q_m=1$. The $N$ coded bits $d_i$, $i=0,1,...,N-1$ with $N=1$, $N=3$ and $N=32$ for $B=1$, $B=2$ and $B>2$, respectively, are given by
\begin{equation}
d_i = 
    \begin{cases}
        b_0 & \text{if $B=1$} \\
        [b_0,b_1,b_2] ~ \text{with $b_2=(b_0+b_1)\mathrm{mod}2$} & \text{if $B=2$} \\
        \left(\sum_{k=0}^{B-1}b_k M_{i,k}\right)\mod{2} & \text{if $B>2$},
    \end{cases}
\end{equation}
where the base sequences $M_{i,k}$ of the Reed-Muller code are defined in \cite[Table 5.3.3.3-1]{ts38212}. 

Subsequently, rate matching is applied to the coded bits $d_i$ to obtain a rate-matched bit sequence $f_k$, $k=0,1,...,E-1$, \cite[Section 5.4.3]{ts38212}
\begin{equation}
    f_k = d_{k~\mathrm{mod}~N}
\end{equation}
with $E=G/2$, where $G=LM$ is the number of OOK symbols available for the LP-WUS.

In the last step, the well-known \textit{Manchester Code} (MC) is applied to $f_k$ resulting in $g_m$, $m=0,1,...,G-1$ coded bits. The rate $R=1/2$ MC maps a single input bit to two coded bits, i.e.
\begin{align}
    [g_{2k},g_{2k+1}] = \begin{cases}
        [1,0] & \text{if } f_k=0 \\
        [0,1] & \text{if } f_k=1.
    \end{cases}
\end{align}
with $k=0,1,...,E-1$. The main advantage of Manchester coding is that it allows for a simple and robust decoder. More precisely, the decoder simply compares a metric (e.g. received energy) corresponding to the two encoded bits as opposed to a threshold which is difficult to obtain in fading channels.

Consider the following example. The gNB configures $N_{SG}^{PO}=7$ subgroups per PO and $N_{PO}^{LO}=1$, i.e. $B=3$ bits to wake up the 7 subgroups as well as all subgroups. For instance, $\mathbf{b}=[0,0,0]$ and $\mathbf{b}=[0,1,1]$ refers to the first subgroup and the fourth subgroup, respectively, whereas $\mathbf{b}=[1,1,1]$ means all 7 subgroups. Furthermore, consider $L=14$ OFDM symbols (1 slot) are available for LP-WUS and $M=2$, hence a total of $G=LM=28$ OOK symbols are available for transmission. The number of bits after channel coding and RM is $E=14$ and after line coding $G$.

% - Encoding
% - Reasons for Manchester coding, interference subtraction etc.
% - ON-sequence Design
% - Encoding of bits to sequences
% - Specific Design choices (ZC,)

\subsubsection{LP-WUS Modulation}
\begin{figure}[htbp]
    \centerline{\includegraphics[width=0.5\textwidth]{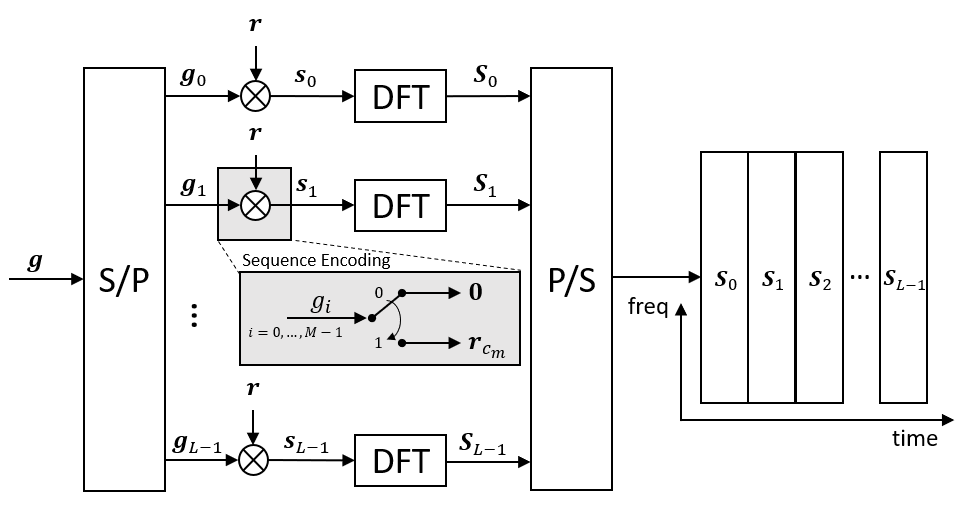}}
    \caption{Block-diagram of LP-WUS modulation.}
    \label{fig:wus-mod}
\end{figure}

The general block diagram of the LP-WUS modulation procedure is depicted in Figure \ref{fig:wus-mod}.

The encoded bits $\mathbf{g}$ are segmented into groups $\mathbf{g}_l=[g_{lM},g_{lM+1},...,g_{(l+1)M-1}]$ of $M$ bits to be modulated onto the $l^{th}$ OFDM symbol configured for LP-WUS. Subsequently, the $i^{th}$ coded bit $g_{li}$ of block $l$ is mapped to an ON-sequence $\mathbf{r} =[r_{0},r_{1},...,r_{\Lon-1}]$ as
\begin{align}
    \mathbf{r} = \begin{cases}
        \mathbf{0}       & \text{if } g_i=0 \\
        \mathbf{r}_{c_m} & \text{if } g_i=1,
    \end{cases}
\end{align}
where $\mathbf{r}_{c_m}$ is the ON-sequence $c_m$, $c_m\in\{0,1,...,N_{seq}-1\}$ denotes the sequence index of the $m^{th}$ OOK ON-symbol ($m=0,1,...,E-1$) and $N_{seq}$ is the number of configured sequences. The WUS time-domain signal $\mathbf{s}_l$ for OFDM symbol $l$ is the concatenation of the $M$ ON-sequences $\mathbf{r}$. 

Subsequently, the time-domain signals $\mathbf{s}_l$ are transformed into frequency-domain via a DFT of size $N_{SC}^{WUS}=132$, and the resulting signals $\mathbf{S}_l$ are allocated to the WUS subcarriers and transmitted on the available WUS time-domain resources.

\subsubsection{Wake-up Sequence Design}

The design of the ON-sequence $\mathbf{r}_{c_m}$, mapped to one OOK symbol of length $\Lon$, is subject to the following criteria: (i) Good spectral properties (e.g. low PAPR), (ii) low specification effort (i.e. reuse of existing sequences) and (iii) good correlation properties.

The sequences in \cite[TS 38.211]{ts38211} are reused and known as cyclically extended Zadoff-Chu sequences, \cite{Jeff}. The $c^{th}$ ON-sequence $\mathbf{r}_c$ reads 
\begin{align}
\label{eq:zc}
    \mathbf{r}_{c}(n) &= \mathbf{x}_q([n+n_{cs}]\mathrm{mod} \Nzc),  \\
    \mathbf{x}_{q}(i) &= e^{-j\frac{\pi q i(i+1)}{\Nzc}} 
\end{align}
with $\Nzc$ denoting the largest prime s.t. $\Nzc<\Lon$, $q\in\{1,2,...,\Nzc-1\}$ is the root and $n_{cs}$ is the cyclic shift of the ZC sequence, $n=0,1,...,\Lon-1$ and $i=0,1,...,\Nzc-1$. Hence ON-sequence $\mathbf{r}_c$ is determined by the root $q$ and a cyclic shift $n_{cs}$.

The number of configurable roots $N_{root}\in\{1,2\}$, which is a compromise between flexibility and the number of correlations that the receiver must perform. Each root requires an additional correlation, whereas sequences with the same root can be detected via the position of the correlation peak corresponding to the cyclic shifts.
To achieve the best correlation performance, the sequences are chosen such that the offset between adjacent cyclic shift values is maximized, i.e.
\begin{equation}
    n_{cs} = (c~\mathrm{mod}~P)\floor*{\frac{\Nzc}{P}} ~,~P=\frac{N_{seq}}{N_{root}},
\end{equation}
where $c=0,1,...,N_{seq}-1$. How the ON-sequences are chosen is described in the next section.

%If only a single sequence is configured, i.e. $N_{seq}=1$, the information bits $\mathbf{b}$ 

%Typically, the possible WUS sequences per OFDM symbol in frequency domain are pre-computed and stored in memory.

% However, given the large amount of potential OOK waveforms which require a significant amount of memory when pre-computed and stored, it is desirable to use a DFT-size which is a power of 2 so that the waveform can be efficiently computed in real-time. Therefore, since the closest power 2 to 132 is 128, the length of the ON-sequence is specified as $\Lon=\{128,64,32\}$. The remaining 4 sub-carriers are left empty and serve as guard-band around the WUS.

\subsubsection{Sequence Encoding}

Multiple sequences can be configured to allow a \textit{coherent} receiver (i.e. with phase information) to obtain the information bits $\mathbf{b}$ through sequence correlation. The maximum number of sequences allowed $N_{seq}^{max}$ depends on $M$ as $N_{seq}^{max}=\{16,8,4\}$ for $M=\{1,2,4\}$, respectively. The reason is that a longer sequence (smaller values of $M$) supports more sequences with good correlation properties than a short sequence. 

The sequences $\mathbf{r}_{c}$ directly encode the information bits $\mathbf{b}$ where the configured number of sequences $N_{seq}\in\{2,4,8,16\}\leq N_{seq}^{max}$ can encode $\delta=\log_2 N_{seq}$ bits per OOK ON-symbol. Therefore, $\lceil B/\delta\rceil$ ON-symbols are required to encode the payload. 
Hence, the coded bits $\mathbf{d}_s=[d_{s,0},d_{s,1},...,d_{s,N_s-1}]$ of length $N_s=B+B_P$ are obtained by prepending $B_P=(-B\mod{\delta})$ zeros, that is, $\mathbf{d}_s=[\mathbf{0}, \mathbf{b}]$. Note that zeros are added \textit{before} the MSB so that both $\mathbf{d}_s$ and $\mathbf{b}$ still encode the same subgroup IDs.
Subsequently, rate-matching is repeating $\mathbf{d}_s$ to obtain $\mathbf{f}_s$ as 
\begin{equation}
    f_{s,i} = d_{s,(i~\mathrm{mod}~N_s)}
\end{equation}
with $i=0,1,...,E_s-1$, where $E_s=E\delta$ and $E=LM/2$ is the number of OOK ON-symbols available. Segment $\mathbf{f}_s$ into blocks of $\delta$ bits such that $\mathbf{f}_s=[\mathbf{f}_{s,0},\mathbf{f}_{s,1},...,\mathbf{f}_{s,(E-1)}]$ where block $m=0,1,...,E-1$ is given by
\begin{equation}
    \mathbf{f}_{s,m} = [f_{s,\delta m},f_{s,\delta m+1},...,f_{s,\delta(m+1)-1}].
\end{equation}
Each block $\mathbf{f}_{s,m}$ is encoded with sequence index $c_m\in\{0,1,...,N_{seq}-1\}$ as
\begin{equation}
    c_m = (\mathbf{f}_{s,m})_{(10)}
\end{equation}
where $(\mathbf{b})_{(10)}$ converts the binary sequence $\mathbf{b}=[b_0,b_1,...]$ to its decimal representation with $b_0$ as MSB. Therefore, the sequence $\mathbf{r}$ of OOK ON-symbol $m$ is given by $\mathbf{r}_{c_m}$.

\begin{figure}[htbp]
    \centerline{\includegraphics[width=0.4\textwidth]{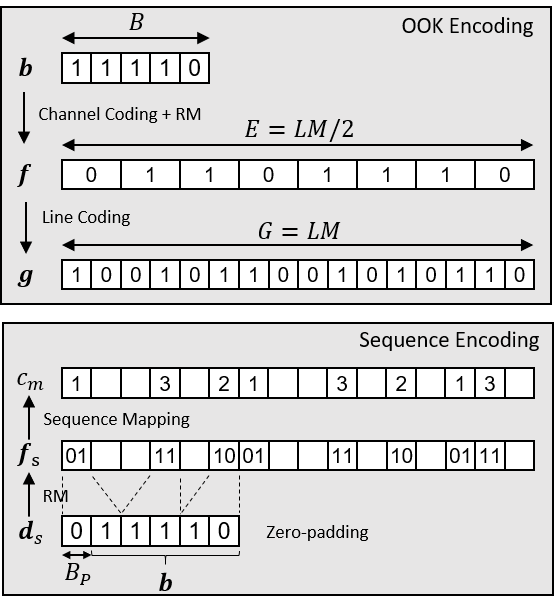}}
    \caption{Example of WUS encoding with $B=5$, $L=4$ and $M=4$.}
    \label{fig:encoding}
\end{figure}

An example of the encoding of payload $\mathbf{b}=[11110]$ is shown in Figure \ref{fig:encoding}.

\subsection{LP-SS Design}
The design of the LP-SS is subject to three main criteria: (i) measurement accuracy, (ii) synchronization accuracy, and (iii) good correlation properties. To mitigate interference from neighboring cells, the network can configure one of four different LP-SS per cell.

%It was agreed to specify 4 LP-SS as a compromise between cross-correlation properties and 

%The network can configure one of 4 LP-SS (a compromise between smaller and larger proposed values) to allow for interference randomization across neighboring cells. 

The LP-SS is a binary sequence $\mathbf{b}_{LPSS}$ of length $B_{LPSS}$ depending on $M_{LPSS}\in\{1,2,4\}$, the configured number of OOK symbols within an OFDM symbol. To achieve sufficient synchronization accuracy for LP-WUS detection, $M_{LPSS}\geq M$. The four LP-SS specified for shorter sequence lengths are given in Table \ref{tab:lpss}.

\begin{table}[ht!]
  \centering
  \begin{tabular}{llll}
    \toprule
    & \multicolumn{3}{c}{$\mathbf{\{B_{LPSS},M_{LPSS},L_{LPSS}\}}$} \\
    \midrule
    $\mathbf{\#}$ & $\{6,1,6\}$ & $\{12,2,6\}$ &  $\{16,4,4\}$ \\
    \midrule
    0 & $[1 0 1 0 1 0]$ & $[1 0 0 1 1 0 0 1 1 0 0 1]$ & $[0 1 1 0 1 0 0 1 1 0 1 0 1 0 1 0]$ \\
    1 & $[0 1 0 1 0 1]$ & $[0 1 1 0 1 0 0 1 1 0 0 1]$ & $[0 1 1 0 1 0 1 0 1 0 0 1 1 0 1 0]$ \\
    2 & $[1 0 0 1 0 1]$ & $[0 1 1 0 0 1 1 0 1 0 0 1]$ & $[1 0 1 0 0 1 1 0 1 0 1 0 1 0 0 1]$ \\
    3 & $[1 0 1 0 0 1]$ & $[0 1 1 0 0 1 0 1 1 0 0 1]$ & $[1 0 1 0 1 0 0 1 1 0 1 0 0 1 1 0]$ \\
    \bottomrule
  \end{tabular}
  \vspace{5pt}
  \caption{Binary LP-SS $\mathbf{b}_{LPSS}$ of the short sequences.}
  \label{tab:lpss}
\end{table}

Note that $\mathbf{b}_{LPSS}$ are balanced, that is, they contain an equal number of ones and zeros, which is beneficial to achieve accurate measurements.

Each bit of $\mathbf{b}_{LPSS}$ is mapped to sequence $\mathbf{0}$ if $b_{LPSS,i}=0$ and $\mathbf{r}$ if $b_{LPSS,i}=1$, where $\mathbf{r}$ is given by \eqref{eq:zc} with $n_{cs}=0$ and root $q$ configured by the network. There is one exception for $M_{LPSS}=1$, where the network is allowed to \textit{not} configure the root $q$ which indicates that an unspecified ON-sequence $\mathbf{r}$ is used. This enables the network to use legacy NR signals on the OOK ON-symbols, e.g. PSS or TRS, which optimizes resource utilization and allows for flexible scheduling. 

\subsubsection{Measurements Metrics}
In this section, we discuss the measurement metrics defined for LP-SS serving cell measurements, \cite{ts38215}. Denote $\mathbf{y}_{LPSS,i}$ the received signal in OOK symbol $i$ when LP-SS is transmitted. Further, define $\mathcal{S}_{ON}$ and $\mathcal{S}_{OFF}$ as the set of OOK ON-and OFF-symbols, respectively, with $|\mathcal{S}_{ON}|=|\mathcal{S}_{OFF}|$ their respective sizes, i.e. the number of corresponding OOK symbols.

\paragraph{LP-RSSI}
LP-RSSI $\Prssi$ is the linear average of the total received power of the LP-SS transmission, i.e.
\begin{equation}
    \Prssi = \frac{1}{B_{LPSS}} \sum_{i=0}^{B_{LPSS}-1} \|\mathbf{y}_{LPSS,i}\|^2
\end{equation}

\paragraph{LP-RSRP} The LP-RSRP $\Prsrp$ is the linear average of the received power of LP-SS in OOK ON-symbols:
\begin{equation}
    \Prsrp = \frac{1}{|\mathcal{S}_{ON}|} \sum_{i\in\mathcal{S}_{ON}} \|\mathbf{y}_{LPSS,i}\|^2.
\end{equation}

\paragraph{LP-RSRQ}
The LP-RSRQ $\Rrsrq$ is the ratio of LP-RSRP and LP-RSSI, that is
\begin{equation}
    \Rrsrq = \frac{\Prsrp}{\Prssi}.
\end{equation}
This ratio provides a measure of interference/noise impact, i.e. the lower the ration, the more noise and interference are present in the received signal. Since both LP-RSRP and LP-RSRQ are measured on the same signal, the LP-RSRQ cannot be greater than 1.

\subsection{Receiver Design}
There are two receiver types considered: (i) an envelope or energy detector (ED) and (ii) a coherent detector (CD). An ED is simpler and consumes less energy than a CD, because it does not have I/Q branches to track both amplitude and phase.
The ED can only decode the OOK-modulated LP-WUS, whereas the CD can correlate with the multiple ON-sequences (if configured) and potentially decode the payload faster. 

Denote the base-band received signal $\mathbf{y}=[\mathbf{y}_0,\mathbf{y}_1,...,\mathbf{y}_{G-1}]$ where $\mathbf{y}_i\in\mathcal{C}^{1\times N}$ is the signal of OOK symbol $i$ containing $N$ complex samples and $G$ is the number of OOK symbols. A potential ED computes the energy $e_i$ for each OOK symbol as
\begin{equation}
    e_i = \|\mathbf{y}_i\|^2 = \mathbf{y}_i\mathbf{y}_i^H.
\end{equation}
Subsequently, the line coding allows to compute an estimate $\hat{\mathbf{f}}=[\hat{f}_0,\hat{f}_1,...,\hat{f}_{G-1}]$ of the channel coded bits as
\begin{align}
    \hat{f}_k = \begin{cases}
        0 & \text{if } e_{2k} > e_{2k+1}\\
        1 & \text{if } e_{2k} \leq e_{2k+1},
    \end{cases}
\end{align}
with $k=0,1,...,G-1$. Alternatively, to improve decoding performance, softbits for $\hat{f}_k$ can be obtained by computing the difference $e_{2k} - e_{2k+1}$. The estimate $\hat{\mathbf{f}}$ is then passed to the channel decoder to obtain the payload estimate $\hat{\mathbf{b}}$. A different implementation can also directly correlate the received envelope with all possible encoded messages resulting in a higher processing gain and improved performance. 

A potential CD correlates the received signal $\mathbf{y}_m$ of OOK ON-symbol $m$ with all possible ON-sequences $\mathbf{r}_c$ and selects the sequence index $\hat{c}_m$ with the highest correlation value, i.e.
\begin{equation}
    \hat{c}_m = \underset{0\leq c < N_{seq}}{\arg\max}\left\{\mathbf{y}_m\mathbf{r}_{c}^H\right\}.
\end{equation}
Subsequently, it is straightforward to obtain a codepoint estimate $\hat{\mathbf{b}}$ from the rate-matched bits $\hat{\mathbf{f}}_{s,m}=(\hat{c}_m)_2$.

% - Basic receiver types, ED and correlation-baed receivers
% - LP-SS mostly for ED, CBR can use SSB
% - signal processing of receivers
% - uncertainty window, multiple windows for decoding
% - sliding window to account for time drift

\subsection{Comparison to IEEE 802.11ba}
The Wi-Fi standard IEEE 802.11ba \cite{b2} also specifies an LP-WUS. Since both 5G and Wi-Fi are based on CP-OFDM, comparing the respective design choices is of particular interest. A first notable difference is that Wi-Fi does not support mobility; therefore, a dedicated signal for low-power measurements such as the LP-SS is not required. In 3GPP, the LP-SS also serves as a synchronization reference for LP-WUS reception, whereas Wi-Fi uses a preamble preceding the WUS for the same purpose. Although a preamble-based solution was considered in 3GPP, it was eventually deemed unnecessary, as synchronization performance can instead be controlled by adjusting the LP-SS periodicity.

Another key difference lies in the subcarrier spacing. Wi-Fi uses an SCS of 312.5 kHz, which is significantly higher than in 3GPP; this results in very short OFDM symbol durations. Consequently, Wi-Fi does not need to multiplex multiple OOK symbols within one OFDM symbol, and therefore uses $M=1$. However, two data rates are supported in Wi-Fi: the higher rate is achieved by using only every second subcarrier, effectively halving the symbol duration. Both systems employ Manchester coding, but only 3GPP applies channel coding to enhance coverage. Additionally, Wi-Fi does not specify the ON-sequence (BPSK and QAM modulation are used), and therefore the LP-WUS in Wi-Fi can be received only with an envelope detector.

\subsection{WUS Design in 6G}
Energy efficiency is one of the key components of 6G and on the UE side, a unified WUS design is under discussion. One major issue with the LP-WUS in 5G is coverage, because a paging message cannot be missed due to non-detection of the WUS. An ED detector has a significantly reduced coverage than a CD and also requires the LP-SS for synchronization and measurements. A WUS design based on CD eliminates these drawbacks since the CD has better performance and can use the legacy SSB for synchronization and measurements. Hence, the current discussions move away from OOK modulation and favor designs based on coherent detection.

\section{Numerical Evaluation}
In this section, we evaluate the performance in terms of BLER, i.e. if the entire WUS payload is received correctly or not, for three different receiver types. The ED accumulating energy per OOK symbol and computing softbits for channel decoding, A CD-S, performing CD through sequence correlation to obtain $B$ in the shortest amount of time and CD-L which correlates over the entire WUS transmission.

% The underlying simulation assumptions are summarized in Table \ref{tab:sim_assumptions}.

\begin{table}[ht!]
  \centering
  \begin{tabular}{ll}
    \toprule
    \textbf{Parameter} & \textbf{Value} \\
    \midrule
    \textbf{Carrier Frequency} & 2.6 GHz \\
    \textbf{Sub-carrier Spacing} & 30 kHz \\
    \textbf{System BW} & 20 MHz (51 PRBs) \\
    \textbf{Antenna Config} & SISO \\
    \textbf{WUS BW} & 5 MHz (132 SCs) \\
    \textbf{WUS Sampling Rate} & 7.68 MHz \\
    \textbf{LP Filter} & 3rd order Butterworth \\
    \textbf{LP Filter BW} & 4.32 MHz \\
    \textbf{Channel Model} & TDL-C 300ns \\
    \bottomrule
  \end{tabular}
  \vspace{5pt}
  \caption{Simulation assumption.}
  \label{tab:sim_assumptions}
  % \vspace{-2mm} 
\end{table}

Figure \ref{fig:tdlc_rx} shows the BLER performance for $B=5$ for the different receivers. Overall, it can be seen that CD performs significantly better than ED. As expected, CD-L has the best performance independent of the number of sequences used. On the other hand, the performance of CD-S depends on $N_{seq}$, i.e. for $N_{seq}=2$, CD-S requires 5 OFDM symbols to recover the payload $B$ whereas for $N_{seq}=8$, only 2 OFDM symbols are required. This results in a 5dB SNR loss at 1\%BLER, because less energy is captured and the cross-correlation among the sequences increased. This highlights the trade-off for CD, shorter detection saves power at the expense of detection performance. 
A practical receiver, will likely be hybrid, i.e. employing ED in good channel conditions and switching to CD if conditions get worse.

\begin{figure}[t]
  \centering
  \begin{tikzpicture}
  \tikzstyle{every pin}=[fill=white,draw=black]
  \pgfplotsset{every axis legend/.append style={
     cells={anchor=west},nodes={scale=0.7, transform shape}}}
  % \pgfplotsset{every axis legend/.append style={
     % cells={anchor=west},nodes={scale=0.7, transform shape}, at={(0.5,1.05)}, anchor=south}}
 %   \pgfplotsset{every axis plot/.append style={smooth}}
    \pgfplotsset{every axis/.append style={line width=0.5pt}}
    \pgfplotsset{every axis/.append style={mark options=solid, mark size=2.5pt}}

    \begin{semilogyaxis}[xlabel={SNR [dB]}, ylabel={BLER},
      grid=minor, xmin=-17, xmax=2, xtick={-17,-15,...,2}, xmajorgrids, ymin=0.001,
      ymax=1,ytickten={-3,-2,-1,0}, legend columns=1]

      % N_seq = 2
      % ED
      \addplot[red, solid, mark=o] plot coordinates {(-10.000000,0.305310) (-9.000000,0.226690) (-8.000000,0.160670) (-7.000000,0.109710) (-6.000000,0.069400) (-5.000000,0.041930) (-4.000000,0.022880) (-3.000000,0.011890) (-2.000000,0.006120) (-1.000000,0.002800) (0.000000,0.001120)};

      % CD for fastest payload recovery
      \addplot[blue, solid, mark=square] plot coordinates {(-17.000000,0.245370) (-16.000000,0.182680) (-15.000000,0.127320) (-14.000000,0.086800) (-13.000000,0.055720) (-12.000000,0.034740) (-11.000000,0.019590) (-10.000000,0.011100) (-9.000000,0.005730) (-8.000000,0.002690) (-7.000000,0.001490) (-6.000000,0.000450) (-5.000000,0.000170) (-4.000000,0.000060) (-3.000000,0.000030) (-2.000000,0.000000) };

      % CD with entire WUS
      \addplot[orange, solid, mark=triangle] plot coordinates {(-17.000000,0.156930) (-16.000000,0.109090) (-15.000000,0.072130) (-14.000000,0.045990) (-13.000000,0.026540) (-12.000000,0.015400) (-11.000000,0.008050) (-10.000000,0.004050) (-9.000000,0.002180) (-8.000000,0.000840) (-7.000000,0.000330) (-6.000000,0.000110) (-5.000000,0.000040) (-4.000000,0.000020) (-3.000000,0.000000) (-2.000000,0.000000)};

      % N_seq = 8
      \addplot[blue, dashed, mark=square] plot coordinates {(-17.000000,0.602310) (-16.000000,0.514520) (-15.000000,0.426200) (-14.000000,0.343890) (-13.000000,0.267990) (-12.000000,0.199780) (-11.000000,0.144840) (-10.000000,0.100900) (-9.000000,0.068190) (-8.000000,0.044180) (-7.000000,0.026790) (-6.000000,0.016960) (-5.000000,0.010030) (-4.000000,0.005610) (-3.000000,0.003530) (-2.000000,0.002160)};

      % CD with entire WUS
      \addplot[orange, dashed, mark=triangle] plot coordinates {(-17.000000,0.157980) (-16.000000,0.108010) (-15.000000,0.072370) (-14.000000,0.045460) (-13.000000,0.027480) (-12.000000,0.015150) (-11.000000,0.008510) (-10.000000,0.003870) (-9.000000,0.001640) (-8.000000,0.000730) (-7.000000,0.000240) (-6.000000,0.000050) (-5.000000,0.000020) (-4.000000,0.000000) (-3.000000,0.000000) (-2.000000,0.000000) };

      \legend{ {ED}\\
               {CD-S, $N_{seq}=2$}\\
               {CD-L, $N_{seq}=2$}\\
               {CD-S, $N_{seq}=8$}\\
               {CD-L, $N_{seq}=8$}\\};

    \end{semilogyaxis}
  \end{tikzpicture}
  \vspace{-10pt}
  \caption{Performance comparison, $B=5$, $M=2$, $L=14$, Table \ref{tab:sim_assumptions}.}
  \label{fig:tdlc_rx}
\end{figure}

\section{Conclusion}\label{sec:conclusion}
This tutorial paper provides a comprehensive overview of the LP-WUS feature for the RRC\_IDLE/INACTIVE states in Release 19, including a detailed discussion of the physical layer design of both LP-WUS and the LP-SS. Energy efficiency is a key objective for 6G, and developing a unified WUS design that achieves an optimal balance between power-saving gains and coverage remains a key challenge to be addressed in 6G.

\vfill

\end{document}